\def\bq{\begin{equation}}
\def\eq{\end{equation}}
\def\bqa{\begin{eqnarray}}
\def\eqa{\end{eqnarray}}
\def\bqb{\begin{eqnarray*}}
\def\eqb{\end{eqnarray*}}
\documentstyle[12pt]{article}\textwidth 16cm
\def\pr#1#2#3{ Phys. Rev. ${\bf{#1}}$ (#2) #3 }
\def\prl#1#2#3{ Phys. Rev. Lett. ${\bf{#1}}$ (#2) #3 }
\def\pl#1#2#3{ Phys. Lett. ${\bf{#1}}$ (#2) #3 }

\def\np#1#2#3{ Nucl. Phys. ${\bf{#1}}$ (#2) #3 }
\def\zp#1#2#3{ Z. Phys. ${\bf{#1}}$ (#2) #3 }

\def\ie{{\it i.e.\/}}
\def\eg{{\it e.g.\/}}
\def\etc{{\it etc.\/}}
\def\etal{{\it et.al.\/}}

\def\L{ {\cal L }}
\def\O{ {\cal O }}

\global\nulldelimiterspace = 0pt

\def\hf{{1\over 2}}

\def\roughly#1{\mathrel{\raise.3ex
    \hbox{$#1$\kern-.75em\lower1ex\hbox{$\sim$}}}}
\def\lsim{\roughly<}
\def\gsim{\roughly>}

\def\sw{s^2_W}

\def\mw{M_W^2}
\def\lw{\lambda_W}
\def\lsm{\biggm( {\lw s\over\mw}\biggm) }
\def\rd{\sqrt2}

\begin{document}
\pagenumbering{arabic}
\thispagestyle{empty}

\begin{flushright} PM/93-37 \\ THES-TP 93/11 \\
October 1993 \end{flushright}
\vspace{2cm}
%---------------------titre ---------------------------------------
\begin{center}
{\Large\bf Unitarity Constraints for Transverse Gauge Bosons
  at LEP and Supercolliders}\dag{~}
 \vspace{1.5cm}  \\
%-----------------------------------------------------------------
 {\Large  G.J. Gounaris } \\
 Department of Theoretical Physics \\
University of Thessaloniki, Greece
\vspace {0.5cm}  \\
 {\Large  J. Layssac, F.M. Renard }\\
 Physique
Math\'{e}matique et Th\'{e}orique,\\
CNRS-URA 768, Universit\'{e} Montpellier II \\
 F-34095 Montpellier Cedex 5

\vspace {2cm}

 {\bf Abstract}
\end{center}

\noindent
Using  asymptotic helicity amplitudes for
Vector-Vector, Vector-Higgs  and
Higgs-Higgs  scattering, we establish  the unitarity constraints on
the $SU(2)_c$
conserving and $W_\mu$ depending interactions, which at sufficiently
high energies may create strong forces among the transverse
vector boson and Higgs states. We then derive
upper bounds for the couplings of these interactions,
which depend on the high energy scale where
unitarity is saturated. If \eg\, unitarity is saturated at 1TeV,
then  $\lambda_W \lsim 0.12$ is obtained.
The implied relations between the present and future LEP
results, and the possible observations of strong effects
 at supercolliders, are also discussed.
\vspace{3cm} \\
\dag{~}Work supported by the scientific cooperation program between
 CNRS
and EIE.
\setcounter{page}{0}

\clearpage

At present the main meagre indication for New Physics(NP)
beyond the Standard Model(SM) comes from the dark matter
problem, the deficit in the
solar neutrino flux \cite{solar} and some peculiarities in double
beta decay spectra \cite{beta}. These may be hinting at some non
vanishing neutrino masses and/or the existence of majorons or
other exotic particles, which clearly belong to NP.
On the other hand, for the gauge boson-fermion interactions
everything looks quite
standard. To some extend, the same is also true for the gauge
boson self-interactions, although the present
precision results mainly from LEP1, still allow
considerable discrepancies from SM. LEP2 and future
supercolliders will look at these gauge boson self interactions
and scrutinize whether there is any NP hidden there.\par

There are many ideas concerning the form of the NP\@.
Inevitably, they all involve new degrees of freedom\footnote{The
usual Higgs doublet is taken as part of SM and not as part of the
new degrees of freedom.}. These degrees of freedom may either be
too heavy to be directly produced at the contemplated
supercolliders, or they might appear as light weakly
interacting particles (like axions, majorons, photinos \etc)
that are difficult to be discovered. In both cases it is
reasonable to expect that at some level they will affect the
gauge boson self interaction, as well as
their interaction with
the standard physical Higgs particle. It should therefore be
fruitfull to look for NP by studying
the gauge boson and Higgs
sector. In fact this may turn out to be the only way
to look for NP in the first of the two cases mentioned above,
where no NP particles can be directly produced at the future
supercolliders. In such a case the appearance of new strong
interactions at sufficiently high energies seems inevitable.
And this is the case that we would like to concentrate on
in the present work.

Even if all new degrees of freedom
are very heavy,  there are still many possibilities for the
form of the high energy strong interactions induced by NP.
In order to reduce them we need
a \underline{principle}. It turns out that such a principle
is available. Motivated by the fact that the
neutral to charge current ratio satisfies $\rho \simeq 1$, it
has been emphasized by many authors that the gauge boson and Higgs
self interactions should be invariant under global $SU(2)_c$
transformations \cite{appel}. Under this assumption, there are
two alternative approaches that have been followed.\par

In the first approach, originally emphasized by Lee Quigg and
Thacker \cite{lqt} and subsequently by others \cite{cg,others},
the additional assumption is made that the $SU(2)_c$
invariant operators induced by NP depend on the scalar fields
\underline{only}. If the couplings of these operators are
sufficiently large, then the NP at the TeV scale appears
in the form of a strong interaction
affecting the longitudinal gauge bosons only, while
the transverse gauge bosons continue to interact weakly.
In this approach, the Higgs particle should be very
heavy; \ \ie \,  $M_H \gsim 1TeV$.\par

In the second alternative suggested in Refs. \cite{gr1,gr2}, no
restriction is imposed on the type of the SM fields appearing in the
$SU(2)_c$ invariant operators induced by NP. Thus the NP operators
are allowed to depend
on the gauge field $W_{\mu}$ also. Restricting their dimensions
to be at most six, there exist only two
such operators called $\O_W$ and $\O_{UW}$. Provided their
couplings are sufficiently large,
strong interactions may be generated
at the TeV scale, albeit among the transverse gauge
bosons now. The mass of the physical Higgs is not affected by
these couplings at tree level. Thus, we may have $M_H\sim M_W$,
so that the longitudinal gauge bosons continue interacting
weakly at all energies. On the
other hand, if it happens that $M_H \gg M_W$ and the coupling of
either $\O_W$ or $\O_{UW}$ is large enough, then the
two alternatives coexist; \ie \, both the transverse
as well as the longitudinal gauge
bosons interact strongly at TeV energies.\par

The unitarity constraints for the
first alternative mentioned above, have been studied long
ago \cite{lqt}. Assuming that the Higgs particle is very heavy,
these constraints give the energy scale at which the
longitudinal gauge boson amplitudes reach unitarity values. Many
authors have studied the signatures of these
strong interactions at various supercollider
experiments \cite{cg,others,bagger}.\par

In the present paper we study the unitarity constraints in the
second alternative approach, where the high energy strong
interactions affect the transverse gauge bosons only.
To this aim we study in detail the $\O_W$ and $\O_{UW}$
contributions to the
vector boson and Higgs scattering amplitudes.
To isolate the second alternative from the first, we assume
$M_H\sim M_W$. As in \cite{lqt,cg} we work
with tree-level amplitudes. We then find that if unitarity is
saturated at the TeV scale or higher, then the couplings of the
operators $\O_W$ and $\O_{UW}$ are
very efficiently constrained. This constraint is in fact so
strong, that if $\O_W$ is indeed observed at LEP2, then
it should be concluded that either strong
interactions must appear at about 1TeV, or
new particles below 1 TeV.\par

More spectacularly, our unitarity constraints appear to be
much more efficient than those obtained from radiative corrections
measurable at LEP1 \cite{deRu}. Of similar nature are also the
unitarity constraints that we find for $\O_{UW}$.
We should also mention that these
constraints are  more
stringent than those obtained
in \cite{Ba} by studying fermion-fermion annihilation to
$W^+W^-$, $WZ$ and $W\gamma$,
even though no $SU(2)\times U(1)$ gauge invariance was imposed
there, which should certainly further weaken them.
Below we derive
our results and subsequently discuss them.\par

It has been shown in \cite{gr2} that, if we impose CP and $SU(2)_c$
symmetries as well as $SU(2)\times U(1)$ gauge
invariance, and we restrict  to operators of dimension up to
six, then the effective lagrangian describing
boson production at the various colliders is completely
described by \\ [.5cm]
\bq \L=\L_{SM}+\L_{NP}  \ \ \ \ \ , \eq

\bqa  \L_{SM}&=& -{1\over 2}\langle W_{\mu\nu}W^{\mu\nu}\rangle-
{1\over4}
  B_{\mu\nu}B^{\mu\nu} +{v^2\over 4}\langle D_\mu
UD^\mu U^{\dagger}\rangle\nonumber\\[.5cm]
 \null & \null & -{v^2 M^2_H\over 8}
\left(\ {1\over 2}\ \langle UU^{\dagger}\rangle-1\right)^2+\
 \makebox{fermionic terms \ \ \ \ , } \eqa

\bq \L_{NP}=\lambda_W{g_2\over M^2_W}\O_W+d\O_{UW} \ \ \ \ , \eq

\noindent where \\
\bq \O_W={1\over3!}\left( \overrightarrow{W}^{\ \ \nu}_\mu\times
  \overrightarrow{W}^{\ \ \lambda}_\nu \right) \cdot
  \overrightarrow{W}^{\ \ \mu}_\lambda =-{2i\over3}
\langle W^{\nu\lambda}W_{\lambda\mu}W^\mu_{\ \ \nu}\rangle \ \ \
, \eq\\
\bq \O_{UW}=\langle (UU^{\dagger}-1)\ W^{\mu\nu} \
W_{\mu\nu}\rangle \ \ \ , \eq
\bq U=\bigm(\widetilde \Phi\ \ , \ \Phi\bigm){\sqrt2\over v} \ \ \ \
. \eq
\noindent Here  $\Phi$ is the standard Higgs
doublet, and the definitions $v^{-2}=\sqrt2\, G_F$, $\widetilde
\Phi = i\tau_2 \Phi^*$
and $\langle A \rangle \equiv TrA$  are  used.\par

Since the operators $\O_W$ and $\O_{UW}$ have a high dimension
($dim=6$), they tend to create strong forces at high
energies. It turns out that these forces affect only
the transverse gauge bosons.
 In order to separate them from the
extensively studied strong forces
among the longitudinal gauge bosons, which are generated
by the scalar
sector in (2) in case the Higgs particle is
heavy \cite{appel,cg,others}, we concentrate here on
the situation of a "light" Higgs;
i.e.\@ $M_H \sim M_W$. In this scenario, only the transverse
gauge bosons may interact strongly at the few TeV scale.\par

The unitarity constraints for $\O_W$ and $\O_{UW}$ come from
studying four body amplitudes involving gauge bosons and the
 physical Higgs as external particles. Convenient approximate
expressions,
valid at the per cent level for CM energies $\sqrt s \gsim
1TeV$, have been recently
found for the gauge boson amplitudes $VV \to VV$,
with $V= W^+,W^-,Z$ or $\gamma$ \cite{glr1}, and the
amplitudes for $VV \to VH$, and $VV \to HH$ involving external
Higgs particles \cite{glr2}. Using this explicit form of the
covariant helicity amplitudes at high energy,
together with the well
known partial wave expansion \cite{jw}, \\
\bq F( \lambda_1 \lambda_2 \rightarrow \lambda_3 \lambda_4)~=~16\pi
\sum_j  \left(j+\hf \right)\ D^{j*}_{\lambda_1-\lambda_2\,,
\,\lambda_3-\lambda_4}(\phi, \theta,0)
\langle \lambda_3 \lambda_4 |T^j | \lambda_1 \lambda_2 \rangle \ \ , \
\eq
\noindent
we obtain the partial wave amplitudes $T^j$.
Note that in the r.h.s.\@ of (7) a factor
of $\rd$ should be inserted, for any initial or final state
channel involving identical particles.
For our unitarity study, it is sufficient to calculate
these amplitudes for the lowest
total angular momentum $j$ only, since
these should be the most
sensitive ones to possible strong interactions.
In (7) $\lambda_i$ denote the helicities, and the
normalization of the partial wave amplitude $T^j$ is the same as in
$2e^{i \delta}sin\delta$, so that the
unitarity limit for elastic processes is given by $|T^j| \leq 2$.
A stronger constraint could be derived by imposing $|Re(T^j)|
\leq 1$, which could be justified if one would know
the analyticity structure of the true amplitudes at
high energy. In order to be conservative, we shall not use
this more
stringent constraint\footnote{Had we used it, then
the r.h.s.\@ of the
inequalities (11-16) below would be smaller by a factor $\rd$.}.
In the following we shall also treat each
effective operator separately and ignore possible, model dependent,
cancellation effects between  $\O_W$ and $\O_{UW}$.
   \par

We first study the unitarity constraints from the process $W^+W^+\to
W^+W^+$, characterized by total charge $Q=2$ in the initial or final
state. For $j=0$, only channels where both $W^+$'s have the
same helicity, feel strongly the $\O_W$ and $\O_{UW}$ interactions.
Identifying \\
\bq |W^+W^+++\rangle\ \to \ \left( \begin{array}{c}1 \\ 0 \end{array}
\right) \ \ \ ,\ \ \ |W^+W^+--\rangle\ \to \ \left( \begin{array}{c}0
\\ 1 \end{array}
\right)\ \ ,\ \
\eq
\noindent
we find \cite{glr1} that the $\O_W$
contribution at $\sqrt s \gsim 1TeV$ is approximately given by\\

\bq T^{j=0}~=~-\, \frac{\alpha}{24\sw}\, \lsm^2 \left( \begin{array}
{cc}
0 & 1\\ 1 & 0
\end{array} \right) \ \ \ , \ \ \
\eq
while the $\O_{UW}$ contribution is \cite{glr1,glr2}\\
\bq T^{j=0}~=~ \frac{\alpha d^2 s}{4 \mw \sw}\, \left( \begin{array}
{cc}
0 & 1\\ 1 & 0
\end{array} \right) \ \ \ . \ \ \
\eq
Using the eigenvalues of (9) and (10) we then find that the unitarity
limits for $\lw$ and $d$ are respectively given by\\
\bq \left| \frac{s\lambda_W}{\mw} \right| \ \lsim
\sqrt{{48\sw}\over\alpha}~ \simeq 39 \ \ \ \ \ \ ,\ \ \
\eq
and\\
\bq \frac{\sqrt s}{M_W} \,|d| \lsim \sqrt{{8\sw}\over\alpha}~\simeq
16 \ \ \ \ \ \ .\ \ \
\eq \par

Before discussing the significance of these results for LEP2 and
the future supercolliders, we first quote the even stronger
unitarity constrains obtained from a similar study of channels
with total charge $Q=1$ and $Q=0$ in the incoming and outgoing states.
These sectors generally involve a larger number of channels, and thus
bigger matrices.

Thus for  $\O_W$, the dominant $j=0$ partial wave amplitudes in the
$Q=1$  sector involve the channels $|W^+Z++\rangle$, $|W^+Z--\rangle$,
$|W^+\gamma ++\rangle$, $|W^+\gamma --\rangle$ \cite{glr1}. The
$T^0$ matrix is 4x4, and its
largest eigenvalue gives\\

\bq \left| \frac{s\lambda_W}{\mw} \right| \ \lsim
\sqrt{{24\sw}\over\alpha}~ \simeq 27.5 \ \ \ \ \ \ .\ \ \
\eq

\noindent

Similarly for the $Q=0$ sector, the relevant channels are
$|W^+W^+++\rangle$, $|W^+W^+--\rangle$, $|ZZ++\rangle$,
$|ZZ--\rangle$, $|Z\gamma ++\rangle$, $|Z\gamma --\rangle$,
 $|\gamma \gamma ++\rangle$,
$|\gamma \gamma --\rangle$ \cite{glr1}, implying an 8x8 T-matrix
whose largest eigenavalue gives\\

\bq \left| \frac{s\lambda_W}{\mw} \right| \ \lsim
\sqrt{{12\sw}\over \alpha}~ \simeq 19 \ \ \ \ \ \ ,\ \ \
\eq\\

\noindent
Finaly from the $j=1$ amplitude we get
\bq \left| \frac{s\lambda_W}{\mw} \right| \ \lsim
\sqrt{{{96\sw}\over
{5 \alpha}}}~ \simeq 25 \ \ \ \ \ \ ,\ \ \
\eq

\noindent
for both the $Q=1$ and $Q=0$ cases. We note for $Q=1$ the
contributing channels are again $|W^+Z++\rangle$, $|W^+Z--\rangle$,
$|W^+\gamma ++\rangle$, $|W^+\gamma --\rangle$; while in the $Q=0$ case
only the two channels $|W^+W^- ++\rangle$ and $|W^+W^- --\rangle$ are
involved \cite{glr1}.\par

We now turn to $\O_{UW}$. For $j=0$ and $Q=1$, the dominant
contributions arise only from the same channels as
in the corresponding $\O_W$ case \cite{glr1,glr2},
and the largest eigenvalue of the relevant $T^0$ matrix gives
again the same result as in (12).
On the other hand for $Q=0$, we need to calculate
an 11x11 $T^0$ matrix involving the same 8 channels as in the
corresponding $\O_W$ case,
and in addition the channels
$|W^+W^-LL\rangle$, $|ZZLL\rangle$ (containing
longitudinal gauge bosons) and $|HH\rangle$.
Using the amplitudes given in \cite{glr1,glr2} we get\\

\bq
 |d|\ \lsim \  17.6\ \frac{\mw}{s}\ + \ 2.43 \ \frac{M_W}{\sqrt s} \
\ \ \ \ . \ \ \ \eq\\
The most stringent constraints coming from eq.(14) and (16) are
plotted in Fig.1.\par

The unitarity constraints given in (11-16)
 involve products of anomalous couplings times the energy
of the processes, in linear or quadratic forms.
The (approximate) equality signs in (11-16)
determine the CM energy $\sqrt s$ where unitarity is first saturated.
They indicate the intuitively obvious result that
 as the  the couplings
$\lambda_W$ or $d$  (of the high dimension interactions)
become smaller, higher energies are needed in
order to reach the unitarity bound. We also note that in
deriving the $\O_W$ constraints it was sufficient
to retain only terms proportional to $(\lambda_W s/\mw)^2$
in the amplitudes,
while in the $\O_{UW}$ case both the linear
and quadratic $d$ contributions  were important.\par

Since the constraints (11-16) are based on the separate
treatment of the operators
$\O_W$ and $\O_{UW}$, they establish  the
conditions under which each of these operators
may be treated perturbatively. If we had
treated instead only appropriate combinations of $\O_W$ and
$\O_{UW}$, then the $\lambda_W\,-\,d$ domains
allowed by the unitarity of the boson-boson amplitudes
would be enlarged, but  only due to strong cancellations between
two large non-perturabative contributions to these
amplitudes. The perturbative treatment of $\O_W$ and $\O_{UW}$
separately,
would not be guaranteed in that case. Thus,
such results could not be used to discuss the
reliability of the perturbative treatment of quantities
like \eg\, the radiative corrections to the LEP1 measurements
(see below), where only one of the above operators can
contribute.\par

Let us now discuss the interpretation of these results.\par

The first remark consists in assuming that the values of the
couplings $\lambda_W$ and $d$ are fixed by some model. For example one
can just assume that they are determined by a characteristic
scale $\Lambda_s$ originating from the dynamics of a strongly
interacting system, so that we may expect \\
\bq {\lambda_W \over \mw} \sim
{}~\frac{\sqrt{4\pi}}{\Lambda_s^2}\ \ \ \ \ \ , \ \ \ \
{d\over M_W} \sim ~ \frac{\sqrt{4\pi}}{\Lambda_s}\ \ \ , \ \  \eq\\
where $\Lambda_s\gsim 1TeV$. Combining (17)  with
(11-16) we get the allowed range for the CM energies of
the various scattering processes, so that they lie below the point
where unitarity effects
show up. This saturation energy is
obviously of the order of $\Lambda_s$. Thus, from \eg
\, our most
stringent bound for $\O_W$ given in (14) we get \\
\bq s ~\lsim~\sqrt{\frac{12\sw}{4\pi\alpha}} \ \Lambda_s^2
{}~\simeq 5.4 \Lambda_s^2 \ \ \ \ \ . \ \ \  \eq \par

The second remark concerns the lowest order contributions
of $\O_W$ or $\O_{UW}$,
to logarithmically divergent one-loop radiative corrections. In
such a case the role of $s$ in (11-16) is played by the
radiative cut-off $\Lambda^2$. Thus, using \eg \, (14), we conclude
that in order the $\O_W$ one-loop radiative correction to be reliable,
the coupling $\lambda_W$ and the cut-off $\Lambda$ should satisfy
the constrain\\
\bq \left| \frac{\Lambda^2 \lambda_W}{\mw} \right| \ \lsim
\sqrt{{12\sw}\over \alpha}~ \simeq 19 \ \ \ \ \ \ ,\ \ \
\eq\\
which for $\Lambda\lsim1TeV$, implies that\\
\bq |\lambda_W|\lsim0.12 \ \ \ \ \ . \ \ \ \ \eq\\
Correspondingly from (16) we would conclude that the one-loop
radiative corrections due to $\O_{UW}$ can only be reliable if\\
\bq
|d|\ \lsim \ 17.6\ \frac{\mw}{\Lambda^2}\ + \ 2.43 \
\frac{M_W}{\Lambda} \lsim 0.3
\ \ \ \ . \ \ \ \eq\\
Below we will return to the implications
of this result for LEP1 measurements.\par

Finally, the third remark  concerns the direct search for the anomalous
couplings $\lambda_W$ and $d$ at (low) energy, i.e.\@ at energies
much smaller than the characteristic scale where unitarity is
saturated. Suppose that such a "low"
energy experiment gives a hint for a certain value for $\lambda_W$ or
$d$. Then the inequalities (11-16) allow us to predict the
lowest energy squared $s_{strong}$ where strong
interactions are expected to appear. Thus using
(14) for $\lambda_W$,  we get\\
\bq s_{strong}~ \simeq ~ 19\ {M^2_W \over|\lambda_W|}\ \ \ \ . \
\ \eq \\
\noindent
Alternatively, the absence of a direct
effect down to the observability limit for $\lambda_W$ in a
low energy experiment, can be transformed into a lower limit
for the unset of the new $\O_W$ induced strong interaction

\bq s_{strong}~ \gsim~ 19\ {M^2_W \over \sup{|\lambda_W|}}\ \ \ \ , \ \
  \eq \\
\noindent
where $\sup{|\lambda_W|}$ is the upper bound of $|\lambda_W|$.
Substituting in (23) the expected or
quoted \underline{upper bounds} for $|\lambda_W |$ that will be
observable at the various colliders, we give in Table 1 the
corresponding \underline{lower bounds} for the threshold
$\sqrt{s_{strong}}$ of the $\O_W$ induced strong
interaction, to which these colliders are sensitive.\\
\begin{center}
\begin{tabular}{|c|c|c|c|} \hline
\multicolumn{4}{|c|}{Table 1: Observability limits for $\lambda_w$ and
$\sqrt{s_{strong}}$}\\[.1cm] \hline
\multicolumn{1}{|c|}{Collider} &
  \multicolumn{1}{|c|}{$|\lambda_W| \lsim$} &
   \multicolumn{1}{|c|}{$\sqrt{s_{strong}} \gsim$} &
     \multicolumn{1}{|c|}{Reference} \\[.1cm] \hline
 LEP2 170GeV & 0.14 & 0.9TeV & \cite{bilenky} \\
 LEP2 230GeV & 0.06 & 1.4TeV & \cite{bilenky}  \\
 NLC 0.5TeV & 0.008 & 4TeV  & \cite{bmt} \\
 NLC 1TeV &  0.002 & 8TeV & \cite{bmt} \\
 SSC/LHC & 0.01 & 3.5TeV & \cite{gr2} \\ \hline
\end{tabular}
\end{center}
\noindent
Corresponding constraints for $\O_{UW}$ are derived using
(16) together with the result that $\O_{UW}$  should be
observable at SSC/LHC
provided $|d| \gsim 0.1$ \cite{glr1}. From this  we conclude
that SSC/LHC is sensitive to the $\O_{UW}$ induced  strong
interaction, provided its
threshold  is at $\sqrt{s_{strong}}\lsim 2.5TeV$.
These results are further illustrated in Fig.1.\par

Precision tests at LEP1 were also used to set bounds on the
anomalous coupling $\lambda_W$ of $\O_W$ \cite{deRu,hag}.
At the Z peak, $\O_W$ contributes only
through 1-loop radiative corrections to the vector boson
self-energies and the vector-fermion-fermion vertices. The use of
effective operators within loops requires some care as soon as the
high energy part of the loop integral is not negligible, and
especially when it is divergent. This was extensively discussed in
\cite{deRu,hag}. The contribution of $\O_W$ to Z peak
observables turns out to
be logarithmically divergent; \ie\ it varies like
$\lambda_W log({\Lambda\over M_W})$ , where $\Lambda$ is the
cut-off. Thus the
bound obtained from Z peak measurements of $\lambda_W$
corresponds to
ignoring all contributions to the loop integral, which are
  associated with
energies higher than $\Lambda$. So a first remark is that one
can appreciate the sensitivity to this unknown sector by varying
$\Lambda$. A second remark stems from the observation that the
LEP1 result for $\Lambda = 1TeV$ and
$M_H = 100 GeV$ \cite{deRu,hag}
\bq    |\lambda_W| \lsim  0.6 \ \ \ \ \ , \ \ \ \ \    \eq
 casts doubts on the reliability of the one-loop
treatment of $\O_W$, to the extend that it allows a violation of
(19,20).\par

The fact that the anomalous  $\O_W$ and $\O_{UW}$ operators,
 even with very
small couplings, lead at high energy to strong interactions among
transverse W states, should always be kept in mind.
It is just a consequence
of the high dimensions
of these operators. In the present work we have gone beyond this
though, by giving the relations
connecting the low energy relics
(\ie\ the anomalous $\lambda_W$ and d
couplings) to the direct manifestations of unitarity saturation
at high energies.
\par

 Such a situation may be compared to the one mentioned in the
introduction, --the first approach to $SU(2)_c$ realization--,
where strong $W_L$ interactions exist but are not seen at low
energy. Several studies of the effect of the anomalous couplings
($\kappa_Z, \kappa_{\gamma}, \delta_Z,...$)
on the $W_L$ properties have been done
in this framework, using the chiral structure of the scalar
sector \cite{cg,bagger}.
It is implicitly assumed in this framework,
that the anomalous couplings have their
origin in the dynamics associated with the mass generation of the W
bosons. \par

In our case the situation may be different. Although the W mass
generation may still be the motivation for modifying the SM,
the New Physics may also affect directly the $W_T$ interactions.
Such a phenomenon has also appeared previously as a consequence
of extending the gauge sector.
In our case it is generated by
$SU(2)_c$ invariance, which  had not been imposed
so strictly before.
In this paper we
established some milestones for the resulting properties.
We have shown that a new type of a strongly
interacting sector affecting \underline{transverse} W states
may be generated.  The magnitude of the anomalous
 $W_T$ couplings observed
in a "low" energy experiment,  is then directly related to
the high energy scale at which strong interactions
should appear. Moreover, we have found that even
 an upper bound on such an anomalous coupling at low energies,
implies a useful lower bound for the scale of the
expected unitarity effects.
These relations turn out to produce remarkable connections
between
LEP2 measurements and supercollider experiments. \par
\vspace{0.5cm}

{\Large \bf Acknowledgements}\vspace{0.5cm}\par
One of us (GJG) wishes to thank the Laboratoire de Physique
 Math\'{e}matique et
Th\'{e}orique de l'Universit\'{e} Montpellier II, for the
hospitality and kind help he received during his stay in
Montpellier.\\

\newpage

\centerline { {\bf Figure Captions }}\par
 Fig.1 Unitarity constraints for $\lambda_W$ and
d couplings. The light grey ($\lambda_W$) and dark grey (d)
domains show the regions where
strong interaction effects should be prominent. We have indicated the
indirect LEP1 bound on $\lambda_W$ and the observability
limits for direct observation expected at various machines.


\newpage

\begin{thebibliography}{99}

\bibitem{solar} R.Davis, D.S.Harmer and K.C.Hoffman,
\prl{20}{1968}{1205}; R.Davis in Proc.\@ 21st Intern.\@ Cosic
Ray Conf.\@, ed.\@ R.J.Protheroe, Vol 12 (Univ.\@ Adelaide,
Australia, 1990).
GALLEX Colab.\@ \pl{B285}{1992}{376}. K.S.Hirata \etal
\pr{D44}{1991}{2241}. SAGE Colab.\@ \prl{67}{1991}{3332}.
\bibitem{beta} M.Moe \etal Nucl.\@ Phys.\@ (Proc.\@ Suppl.\@)
{\bf B31} (1993); F.T. Avignone \etal \pl{B256}{1991}{559}.
\bibitem{appel} T.Appelquist in "Gauge Theories and Experiments
at High Energies", ed.\@ by K.C. Brower and D.G. Sutherland,
Scottish University Summer School in Physics, St.\@ Andrews
(1980).
\bibitem{lqt} B.W.Lee, C.Quigg and H.Thacker, \pr{D16}{1977}{1519}.
\bibitem{cg} M.S.Chanowitz and M.K.Gaillard, \np{B261}{1985}{379};
    M.S.Chanowitz, Ann.Rev.Nucl.Part.Sci.38(1988)323.
\bibitem{others} Ken-ichi Hikasa, Inv.\@ talk at Workshop on
Physics with Linear Colliders, Finland (1991).
See also A.Dobado, M.J.Herrero and J.Terron,
\zp{C50}{1991}{205}, \zp{C50}{1991}{465} and refences therein.
\bibitem{gr1} G.J.Gounaris and F.M.Renard, \zp{C59}{1993}{133}.
\bibitem{gr2} G.J.Gounaris and F.M.Renard, \zp{C59}{1993}{143}.
\bibitem{bagger} J.Bagger, S.Dawson and G.Valencia,
\np{B399}{1993}{364}.\\
   A.Falk, M.Luke and E.Simmons, \np{B365}{1991}{523}.
\bibitem{deRu} A.De Rujula et al, \np{B384}{1992}{3}.
\bibitem{Ba} U.Baur and D. Zeppenfeld, \pl{B201}{1988}{383}.
\bibitem{glr1} G.J.Gounaris, J.Layssac and F.M.Renard, preprint
   PM/93-26, THES-TP 93/8, to be published.
\bibitem{glr2} G.J.Gounaris J. Layssac and F.M.Renard, in preparation.
\bibitem{jw} M.Jacob and G.C.Wick, Ann.Phys. $\bf{7}$ (1959) 404.
\bibitem{bilenky} M.Bilenky, J.L.Kneur, F.M. Renard and D.
  Schilknecht,\np{B409}{1993}{22}.
\bibitem{bmt}G.J.Gounaris et al, in Proc. of the Workshop on
    $e^+e^-$ Collisions a  500 GeV: The Physics Potential,
    DESY 92-123B(1992), p.735, ed. P.Zerwas.
    M.Bilenky et al, BI-TP in preparation.
\bibitem{hag} K.Hagiwara et al, \pl{B283}{1992}{353},
          \pr{D48}{1993}{2182}.\\
 T.Appelquist and G-H.Wu \pr{D48}{1993}{3235}.\\



\end{thebibliography}
\end{document}